\documentclass[manuscript]{emulateapj}
\usepackage{natbib,amsmath}
\shorttitle{Towards reliable flare forecast benchmarking}
\shortauthors{Bloomfield et al.}

\begin{document}

\title{Towards Reliable Benchmarking of Solar Flare Forecasting Methods}


\author{D. Shaun Bloomfield\altaffilmark{1}, Paul A. Higgins\altaffilmark{1}, R. T. James McAteer\altaffilmark{2},  and Peter T. Gallagher\altaffilmark{1}}
\email{shaun.bloomfield@tcd.ie}
\altaffiltext{1}{Astrophysics Research Group, School of Physics, Trinity College Dublin, College Green, Dublin 2, Ireland.}
\altaffiltext{2}{Department of Astronomy, New Mexico State University, Las Cruces, New Mexico 88003-8001, USA.}

%
%
%
%

\begin{abstract}
Solar flares occur in complex sunspot groups, but it remains unclear how the probability of producing a flare of a given magnitude relates to the characteristics of the sunspot group. Here, we use \emph{Geostationary Operational Environment Satellite} X-ray flares and McIntosh group classifications from solar cycles 21 and 22 to calculate average flare rates for each McIntosh class and use these to determine Poisson probabilities for different flare magnitudes. Forecast verification measures are studied to find optimum thresholds to convert Poisson flare probabilities into yes/no predictions of cycle 23 flares. A case is presented to adopt the true skill statistic (TSS) as a standard for forecast comparison over the commonly used Heidke skill score (HSS). In predicting flares over 24\,hr, the maximum values of TSS achieved are 0.44 (C-class), 0.53 (M-class), 0.74 (X-class), 0.54 ($\geqslant$M1.0), and 0.46 ($\geqslant$C1.0). The maximum values of HSS are 0.38 (C-class), 0.27 (M-class), 0.14 (X-class), 0.28 ($\geqslant$M1.0), and 0.41 ($\geqslant$C1.0). These show that Poisson probabilities perform comparably to some more complex prediction systems, but the overall inaccuracy highlights the problem with using average values to represent flaring rate distributions.
\end{abstract}


\section{Introduction}
\label{sec:int}
Solar flares result from the release of enormous quantities of energy \citep[up to $\sim$10$^{27}$\,J;][]{Kane:2005} from twisted, non-potential magnetic fields. Along with coronal mass ejections (CMEs), flares are a major contributor to space weather that adversely affects the near-Earth environment \citep{Hapgood:2010}. The magnetic energy to power solar flares is stored primarily in active regions (ARs) that are routinely classified in terms of complexity. The Mount Wilson scheme \citep{Hale:1919p44, Kunzel:1960p521} describes magnetic polarity mixing, while the \citet{McIntosh:1990p528} scheme describes spatial structuring of the magnetic field ``footprints'' in sunspot groups. We concentrate on the McIntosh scheme that allows up to 60 classes, yielding reasonable resolution in terms of the observed structural complexity. In contrast, the Mount Wilson scheme allows up to eight classes, each with flare rate distributions more broad than the McIntosh classes.

Recent years have seen a resurgence in the field of solar flare prediction. A sample of the techniques employed includes Poisson statistics \citep{Gallagher:2002p342}, Bayesian statistics \citep{Wheatland:2005p71}, support vector machines \citep{Li:2007p721}, discriminant analysis \citep{Barnes:2007p720}, ordinal logistic regression \citep{Song:2009p703, Yuan:2010p716}, neural networks \citep{Colak:2009p714, Yu:2009p49, Ahmed:2012}, wavelet predictors \citep{Yu:2010p100}, Bayesian networks \citep{Yu:2010p221}, predictor teams \citep{Huang:2010p717}, superposed epoch analysis \citep{Mason:2010p532}, and empirical projections \citep{Falconer:2011}. It is worth noting that none of these techniques are based on physical models of the flare process. Most of the methods give a probability for an X-ray flare with peak flux above some magnitude in a time interval. If the aim of a prediction method is to provide a result that can be readily interpreted as ``flare imminent'' or ``no flare expected'', the predicted probabilities need to be converted into yes/no forecasts and the forecast success determined. However, it is extremely important that appropriate performance measures are used when comparing the success of different forecasts.

In this Letter, a case is presented for the adoption of an existing (but rarely utilized) performance measure for comparisons between different solar flare forecasts (Section~\ref{sec:mea}). As an example, we investigate the performance of Poisson probabilities in predicting X-ray flares from ARs within 24\,hr of a McIntosh classification being issued. The data and their sources are detailed in Section~\ref{sec:dat}, while the method to determine forecast performance is described in Section~\ref{sec:ana_met}. The effect of varying the threshold that is used in converting Poisson probabilities into yes/no predictions is studied in Section~\ref{subsec:thr_var}, while optimum performance measures are compared to the performance of other methods in Section~\ref{subsec:opt_ss_com}. Finally, our conclusions and ideas for further work are given in Section~\ref{sec:con}.

\section{Forecast Performance Measures}
\label{sec:mea}
The success of a forecast method that provides yes/no forecasts should be studied using a forecast contingency table and calculating verification measures \citep[an excellent comparison of different evaluation measures is given in][]{Woodcock:1976p384}. Quantitative measures are essential to compare the relative performance of different prediction methods. The flare forecast contingency table format is presented in Table~\ref{tab:ex_con}, containing the elements TP (true positives, ``flare'' predicted and observed), FN (false negatives, ``no flare'' predicted and flare observed), FP (false positives, ``flare'' predicted and none observed), and TN (true negatives, ``no flare'' predicted and none observed). Numerous skill scores exist to quantify the performance of forecasts, but the \citet{Heidke:1926} skill score (HSS),
\begin{align}
\label{eqn:hss}
\mathrm{HSS}&=\\
&\frac{2[(\mathrm{TP}\times\mathrm{TN})-(\mathrm{FN}\times\mathrm{FP})]}{(\mathrm{TP}+\mathrm{FN})(\mathrm{FN}+\mathrm{TN})+(\mathrm{TP}+\mathrm{FP})(\mathrm{FP}+\mathrm{TN})}\nonumber\ ,
\end{align}
is most frequently used in flare forecasting \citep[e.g.,][]{Barnes:2008p206}. The strength of the HSS lies in its use of the whole contingency table to quantify the accuracy of achieving correct predictions relative to random chance. The \citet{Hanssen:1965} discriminant, known as the true skill statistic (TSS), also uses all of the elements,
\begin{equation}
\label{eqn:tss}
\mathrm{TSS}=\frac{\mathrm{TP}}{\mathrm{TP}+\mathrm{FN}}-\frac{\mathrm{FP}}{\mathrm{FP}+\mathrm{TN}}\ .
\end{equation}
However, only TSS is unbiased when confronted with varying event/no-event sample ratios \citep{Woodcock:1976p384}. This is demonstrated by considering a new forecast that achieves the same prediction success with two times the number of flare ARs (i.e., $\mathrm{TP}_{\mathrm{new}}=2\mathrm{TP}$; $\mathrm{FN}_{\mathrm{new}}=2\mathrm{FN}$; $\mathrm{TP}_{\mathrm{new}}/\mathrm{FN}_{\mathrm{new}}=\mathrm{TP}/\mathrm{FN}$). Equation~\ref{eqn:hss} becomes,
\begin{align}
\label{eqn:hss_new}
&\mathrm{HSS}_\mathrm{new}\nonumber\\
&=\frac{2[(2\mathrm{TP}\times\mathrm{TN})-(2\mathrm{FN}\times\mathrm{FP})]}{(2\mathrm{TP}+2\mathrm{FN})(2\mathrm{FN}+\mathrm{TN})+(2\mathrm{TP}+\mathrm{FP})(\mathrm{FP}+\mathrm{TN})}\nonumber\\
&\neq\mathrm{HSS}\ ,
\end{align}
while Equation~\ref{eqn:tss} becomes,
\begin{align}
\label{eqn:tss_new}
\mathrm{TSS}_\mathrm{new}&=\frac{2\mathrm{TP}}{2\mathrm{TP}+2\mathrm{FN}}-\frac{\mathrm{FP}}{\mathrm{FP}+\mathrm{TN}}\nonumber\ \\
&=\frac{\mathrm{TP}}{\mathrm{TP}+\mathrm{FN}}-\frac{\mathrm{FP}}{\mathrm{FP}+\mathrm{TN}}=\mathrm{TSS}\ .
\end{align}
This simple example shows that HSS changes despite the prediction success being held constant, highlighting the problem with using HSS to compare between different methods (or different trials of the same method). Note that we do not dismiss the usefulness of HSS as a measure within a particular forecast method trial. However, we propose TSS to be the standard measure for comparing between flare forecasts, given that different studies use differing flare/no-flare sample ratios.

\begin{deluxetable}{lcc}
\tabletypesize{\scriptsize}
\tablecaption{Flare Forecast Contingency Table\label{tab:ex_con}}
\tablewidth{\columnwidth}
\tablehead{
\colhead{Flare}		&  \multicolumn{2}{c}{Forecast}\\
\colhead{Observed}	&  \colhead{``Flare''}	&  \colhead{``No flare''}
}
\startdata
Yes				&  TP			&  FN\\
No				&  FP			&  TN
\enddata
\end{deluxetable}

\section{Data Sources}
\label{sec:dat}

\subsection{Training Set}
\label{subsec:training}
In order to facilitate the calculation of flare probabilities, we obtained historical flare rates for each McIntosh class from two locations that share the same data source. The National Oceanic and Atmospheric Administration (NOAA) Space Weather Prediction Center (SWPC) provided total numbers of \emph{Geostationary Operational Environmental Satellite} (\emph{GOES}) C-, M-, and X-class flares and the originating ARs for each McIntosh classification over 1988 December 1 to 1996 June 30 (C.C. Balch 2011, private communication). Additional M- and X-class flare and McIntosh class numbers were taken from \citet{Kildahl:1980p166} over 1969--1976, but relate to the same data source (i.e., NOAA-collated ground-based AR observations and \emph{GOES} flare events). These were included to increase the rare M- and X-class samples so that the rates were more statistically significant. Table~\ref{tab:classes} presents the recorded McIntosh classes with the numbers of observed regions and flares produced.

\begin{deluxetable*}{lrrrrrrrrrrrrrrrrr}[!h]
\tabletypesize{\scriptsize}
\setlength{\tabcolsep}{0.04in} 
\tablecaption{McIntosh Classification Flare Statistics\label{tab:classes}}
\tablewidth{\textwidth}
\tablehead{
\colhead{McIntosh}  &  \multicolumn{4}{c}{SWPC (1988--1996)}  &  \multicolumn{4}{c}{Kildahl (1969--1976)\tablenotemark{b}}  &  \multicolumn{4}{c}{Combined Flare Rate (24\,hr$^{-1}$)}  &  \multicolumn{5}{c}{Poisson Flare Probability (\%)} \\
\colhead{Region}  &  \colhead{Region}  &  \multicolumn{3}{c}{Total Flares}  &  \colhead{Region}  &  \multicolumn{3}{c}{Total Flares}  &  \multicolumn{4}{c}{In \emph{GOES} Class}  &  \multicolumn{3}{c}{In \emph{GOES} Class}  &  \multicolumn{2}{c}{Above \emph{GOES}\tablenotemark{d}} \\
\colhead{Classes\tablenotemark{a}}  &  \colhead{Count}  &  \colhead{C}  &  \colhead{M}  &  \colhead{X}  &  \colhead{Count}  &  \colhead{C\tablenotemark{c}}  &  \colhead{M}  &  \colhead{X}  &  \colhead{C}  &  \colhead{M}  &  \colhead{X}  &  \colhead{$\pm\sigma$}  &  \colhead{C}  &  \colhead{M}  &  \colhead{X}  &  \colhead{M1.0}  &  \colhead{C1.0}  
}
\startdata
AXX  &  2748  &    82  &   10  &   0  &  2517  &   75.1  &   31  &   3  &  0.03  &  0.01  &  0.00  &  0.01  &    3  &    1  &    0  &    1  &    4  \\
BXO  &  3342  &   217  &   18  &   1  &  1906  &  123.8  &   41  &   2  &  0.06  &  0.01  &  0.00  &  0.01  &    6  &    1  &    0  &    1  &    7  \\
BXI  &     0  &     0  &    0  &   0  &   334  &    0.0  &   20  &   0  &  0.00  &  0.06  &  0.00  &  0.05  &    0  &    6  &    0  &    6  &    6  \\
HRX  &   336  &    21  &    1  &   0  &   211  &   13.2  &    7  &   1  &  0.06  &  0.01  &  0.00  &  0.04  &    6  &    1  &    0  &    2  &    8  \\
HSX  &  1968  &    94  &   21  &   0  &  1963  &   93.8  &   99  &   6  &  0.05  &  0.03  &  0.00  &  0.02  &    5  &    3  &    0  &    3  &    8  \\
HAX  &   598  &    49  &   13  &   0  &   222  &   18.2  &   14  &   0  &  0.08  &  0.03  &  0.00  &  0.03  &    8  &    3  &    0  &    3  &   11  \\
HHX  &    53  &     3  &    1  &   0  &   150  &    8.5  &   16  &   2  &  0.06  &  0.08  &  0.01  &  0.07  &    6  &    8  &    1  &    9  &   14  \\
HKX  &    49  &    11  &    2  &   0  &    38  &    8.5  &    7  &   0  &  0.22  &  0.10  &  0.00  &  0.11  &   20  &   10  &    0  &   10  &   28  \\
\tableline
CRO  &   745  &   102  &    3  &   0  &   368  &   50.4  &   20  &   2  &  0.14  &  0.02  &  0.00  &  0.03  &   13  &    2  &    0  &    2  &   15  \\
CRI  &     6  &     2  &    0  &   0  &   152  &   50.7  &    7  &   0  &  0.33  &  0.04  &  0.00  &  0.08  &   28  &    4  &    0  &    4  &   31  \\
CSO  &  1504  &   284  &   27  &   0  &  1020  &  192.6  &   40  &   1  &  0.19  &  0.03  &  0.00  &  0.02  &   17  &    3  &    0  &    3  &   19  \\
CSI  &    14  &     8  &    2  &   0  &   211  &  120.6  &   16  &   2  &  0.57  &  0.08  &  0.01  &  0.07  &   44  &    8  &    1  &    9  &   48  \\
CAO  &  1455  &   361  &   38  &   2  &   232  &   57.6  &   18  &   1  &  0.25  &  0.03  &  0.00  &  0.02  &   22  &    3  &    0  &    3  &   25  \\
CAI  &    27  &    14  &    6  &   0  &   166  &   86.1  &   19  &   0  &  0.52  &  0.13  &  0.00  &  0.07  &   40  &   12  &    0  &   12  &   48  \\
CHO  &    88  &    21  &    2  &   1  &   112  &   26.7  &    8  &   1  &  0.24  &  0.05  &  0.01  &  0.07  &   21  &    5  &    1  &    6  &   26  \\
CHI  &     2  &     1  &    0  &   0  &    29  &   14.5  &    6  &   0  &  0.50  &  0.19  &  0.00  &  0.18  &   39  &   18  &    0  &   18  &   50  \\
CKO  &   135  &    59  &   11  &   0  &    52  &   22.7  &   13  &   2  &  0.44  &  0.13  &  0.01  &  0.07  &   35  &   12  &    1  &   13  &   44  \\
CKI  &    17  &    14  &    6  &   0  &    28  &   23.1  &    6  &   2  &  0.82  &  0.27  &  0.04  &  0.15  &   56  &   23  &    4  &   27  &   68  \\
\tableline
DRO  &    63  &    12  &    3  &   0  &    75  &   14.3  &    6  &   0  &  0.19  &  0.07  &  0.00  &  0.09  &   17  &    6  &    0  &    6  &   23  \\
DRI  &     2  &     7  &    0  &   0  &    54  &  189.0  &    7  &   1  &  3.50  &  0.12  &  0.02  &  0.13  &   97  &   12  &    2  &   13  &   97  \\
DSO  &   546  &   198  &   26  &   1  &   553  &  200.5  &   51  &   6  &  0.36  &  0.07  &  0.01  &  0.03  &   30  &    7  &    1  &    7  &   36  \\
DSI  &    39  &    34  &    6  &   0  &   246  &  214.5  &   31  &   1  &  0.87  &  0.13  &  0.00  &  0.06  &   58  &   12  &    0  &   12  &   63  \\
DSC  &     0  &     0  &    0  &   0  &    20  &    0.0  &    5  &   2  &  0.00  &  0.25  &  0.10  &  0.22  &    0  &   22  &   10  &   30  &   30  \\
DAO  &  1775  &   784  &  124  &   4  &   288  &  127.2  &   28  &   2  &  0.44  &  0.07  &  0.00  &  0.02  &   36  &    7  &    0  &    7  &   40  \\
DAI  &   391  &   419  &   70  &   6  &   324  &  347.2  &   58  &   7  &  1.07  &  0.18  &  0.02  &  0.04  &   66  &   16  &    2  &   18  &   72  \\
DAC  &     8  &     5  &    3  &   0  &    46  &   28.8  &   12  &   1  &  0.62  &  0.28  &  0.02  &  0.14  &   46  &   24  &    2  &   26  &   60  \\
DHO  &    46  &    26  &    1  &   1  &    43  &   24.3  &   11  &   0  &  0.57  &  0.13  &  0.01  &  0.11  &   43  &   13  &    1  &   14  &   51  \\
DHI  &    11  &    14  &    1  &   0  &    41  &   52.2  &    3  &   0  &  1.27  &  0.08  &  0.00  &  0.14  &   72  &    7  &    0  &    7  &   74  \\
DHC  &     0  &     0  &    0  &   0  &     6  &    0.0  &    2  &   0  &  0.00  &  0.33  &  0.00  &  0.41  &    0  &   28  &    0  &   28  &   28  \\
DKO  &   217  &   178  &   55  &   5  &    43  &   35.3  &   14  &   2  &  0.82  &  0.27  &  0.03  &  0.06  &   56  &   23  &    3  &   25  &   67  \\
DKI  &   223  &   288  &   69  &   6  &    88  &  113.7  &   42  &   6  &  1.29  &  0.36  &  0.04  &  0.06  &   73  &   30  &    4  &   33  &   81  \\
DKC  &    57  &    93  &   35  &   5  &   100  &  163.2  &   72  &  10  &  1.63  &  0.68  &  0.10  &  0.08  &   80  &   49  &    9  &   54  &   91  \\
\tableline
ESO  &    95  &    37  &    6  &   0  &    82  &   31.9  &   14  &   0  &  0.39  &  0.11  &  0.00  &  0.08  &   32  &   11  &    0  &   11  &   39  \\
ESI  &    18  &    33  &    1  &   0  &    78  &  143.0  &   22  &   2  &  1.83  &  0.24  &  0.02  &  0.10  &   84  &   21  &    2  &   23  &   88  \\
EAO  &   459  &   267  &   61  &   0  &    47  &   27.3  &   10  &   4  &  0.58  &  0.14  &  0.01  &  0.04  &   44  &   13  &    1  &   14  &   52  \\
EAI  &   295  &   370  &   83  &   2  &    82  &  102.8  &   48  &   1  &  1.25  &  0.35  &  0.01  &  0.05  &   71  &   29  &    1  &   30  &   80  \\
EAC  &     3  &     5  &    1  &   0  &    17  &   28.3  &    6  &   3  &  1.67  &  0.35  &  0.15  &  0.22  &   81  &   30  &   14  &   39  &   89  \\
EHO  &    42  &    31  &    6  &   0  &    39  &   28.8  &    6  &   0  &  0.74  &  0.15  &  0.00  &  0.11  &   52  &   14  &    0  &   14  &   59  \\
EHI  &    15  &    24  &    6  &   0  &    45  &   72.0  &   28  &   4  &  1.60  &  0.57  &  0.07  &  0.13  &   80  &   43  &    6  &   47  &   89  \\
EHC  &     2  &     9  &    0  &   0  &     4  &   18.0  &    8  &   0  &  4.50  &  1.33  &  0.00  &  0.41  &   99  &   74  &    0  &   74  &  100  \\
EKO  &   185  &   173  &   35  &   3  &    52  &   48.6  &   20  &   1  &  0.94  &  0.23  &  0.02  &  0.06  &   61  &   21  &    2  &   22  &   69  \\
EKI  &   423  &   703  &  173  &  23  &    81  &  134.6  &  103  &  11  &  1.66  &  0.55  &  0.07  &  0.04  &   81  &   42  &    7  &   46  &   90  \\
EKC  &   103  &   278  &  132  &  17  &    63  &  170.0  &  149  &  21  &  2.70  &  1.69  &  0.23  &  0.08  &   93  &   82  &   20  &   85  &   99  \\
\tableline
FRI  &     0  &     0  &    0  &   0  &     2  &    0.0  &    1  &   0  &  0.00  &  0.50  &  0.00  &  0.71  &    0  &   39  &    0  &   39  &   39  \\
FSO  &    14  &     9  &    3  &   0  &    13  &    8.4  &    6  &   1  &  0.64  &  0.33  &  0.04  &  0.19  &   47  &   28  &    4  &   31  &   64  \\
FSI  &     6  &    12  &    0  &   0  &     8  &   16.0  &   15  &   0  &  2.00  &  1.07  &  0.00  &  0.27  &   86  &   66  &    0  &   66  &   95  \\
FAO  &    73  &    63  &   16  &   0  &     3  &    2.6  &    0  &   0  &  0.86  &  0.21  &  0.00  &  0.11  &   58  &   19  &    0  &   19  &   66  \\
FAI  &    91  &   106  &   35  &   3  &    12  &   14.0  &    8  &   0  &  1.16  &  0.42  &  0.03  &  0.10  &   69  &   34  &    3  &   36  &   80  \\
FHO  &     9  &     5  &    1  &   0  &    10  &    5.6  &    0  &   0  &  0.56  &  0.05  &  0.00  &  0.23  &   43  &    5  &    0  &    5  &   46  \\
FHI  &    10  &    17  &    9  &   0  &    18  &   30.6  &   15  &   0  &  1.70  &  0.86  &  0.00  &  0.19  &   82  &   58  &    0  &   58  &   92  \\
FHC  &     0  &     0  &    0  &   0  &     5  &    0.0  &    4  &   0  &  0.00  &  0.80  &  0.00  &  0.45  &    0  &   55  &    0  &   55  &   55  \\
FKO  &    97  &   165  &   29  &   1  &    19  &   32.3  &    6  &   0  &  1.70  &  0.30  &  0.01  &  0.09  &   82  &   26  &    1  &   27  &   87  \\
FKI  &   235  &   517  &  161  &  17  &    47  &  103.4  &  106  &  17  &  2.20  &  0.95  &  0.12  &  0.06  &   89  &   61  &   11  &   66  &   96  \\
FKC  &    93  &   233  &  146  &  24  &    27  &   67.6  &   39  &  13  &  2.51  &  1.54  &  0.31  &  0.09  &   92  &   79  &   27  &   84  &   99  
\enddata
\tablenotetext{a}{Only includes classifications producing $\geqslant$1 C-, M-, or X-class flare in either time range.}
\tablenotetext{b}{From \citet{Kildahl:1980p166}.}
\tablenotetext{c}{Non-integer flare numbers result from use of observed C-class rates from SWPC (1988--1996).}
\tablenotetext{d}{``Above \emph{GOES} X1.0'' is equivalent to ``In \emph{GOES} Class X''.}
\end{deluxetable*}

\subsection{Testing Set}
\label{subsec:testing}
The AR and flare data that are used for testing were gathered from the online archives of NOAA/SWPC.\footnote{\url{http://www.swpc.noaa.gov/ftpdir/warehouse/}} McIntosh classes of regions that have predictions issued and tested were taken from the daily NOAA Solar Region Summary files over 1996 August 1 to 2010 December 31. In this work, each daily record of a NOAA region was treated as an individual measurement, yielding 22276 AR samples. \emph{GOES} flares with originating NOAA numbers assigned to their entry were extracted from the edited daily NOAA Solar Event Reports over the same date range as the McIntosh classes. NOAA region numbers attributed to any associated H$\alpha$ flares were used for those \emph{GOES} flares with no NOAA region directly assigned.

\section{Analysis Method}
\label{sec:ana_met}

\subsection{Historical Poisson Probabilities}
\label{subsec:pois_prob}
Following \citet{Bornmann:1994p709}, \emph{GOES}-class flare rates in 24\,hr intervals were calculated for each McIntosh class by combining the number of flares that classification produced over 1969--1976 and 1988--1996 and dividing by the number of times the McIntosh class was observed in both periods, $N_\mathrm{tot}$. It should be noted that C-class flares were not provided in \citet{Kildahl:1980p166}. In order to provide C-class related forecasts comparable to those for M- and X-classes, rates measured over 1988--1996 were taken to hold for 1969--1976. The relative numbers of McIntosh observations in the time periods was then used to determine the expected number of C-class flares for 1969--1976 (Table~\ref{tab:classes}, Column 7). The C-, M-, and X-class flare rates combined over 1969--1976 and 1988--1996 are presented in Columns 10--12 of Table~\ref{tab:classes}, with the error on the average rate ($\sigma=N_\mathrm{tot}^{-1/2}$) given in Column 13.

To achieve a probability of flaring we follow the Poisson statistics technique of \citet{Gallagher:2002p342}. Under the assumption of flares being a Poisson-distributed process,\footnote{\citet{Aschwanden:2010p457} show that flare waiting times are consistent with a nonstationary Poisson process. Application of Poisson probability here averages the time-dependent rates in 24\,hr intervals and over the solar cycle.} the probability of observing $N$ flares in a time interval is related to the average flare rate, $\mu$, over that interval by,
\begin{equation}
P_\mu(N)=\frac{\mu^N}{N!}\exp(-\mu)\ .
\end{equation}
When $\mu$ is calculated over 24\,hr intervals, the probability of observing one or more flares in any 24\,hr interval is,
\begin{eqnarray}
P_\mu(N\geqslant1)&=&1-P_\mu(N=0)\nonumber\ ,\\
				&=&1-\exp(-\mu)\ .
\end{eqnarray}
Poisson probabilities for a McIntosh class to produce at least one flare within a 24\,hr interval are displayed in Columns 14--16 of Table~\ref{tab:classes} for the C-, M-, and X-classes, with those for flaring $\geqslant$M1.0 (M- and X-classes) and $\geqslant$C1.0 (C-, M-, and X-classes) in Columns 17--18.

\subsection{Contingency Table Construction}
\label{subsec:con_tab_ver}
Two sets of binary (yes/no) information are required to build the forecast contingency tables---flare truth and flare prediction. The first is achieved by cross-referencing the SWPC-extracted AR and \emph{GOES} event lists over the testing period (1996--2010). For each AR observed each day, the list of AR-associated flares within 24\,hr of the McIntosh class being issued is searched for the NOAA number of that AR (i.e., the same UT day; McIntosh classes are published at 00:30\,UT based on data before 00:00\,UT). Flare truth is set to ``no'' for ARs when no flares occurred with peak magnitude at the appropriate level or ``yes'' when $\geqslant$1 flare occurred. This results in the number of flare ARs, $N_{\mathrm{fl}}$, being 3667, 810, and 92 for C-, M-, and X-class events, respectively. Similarly, $N_{\mathrm{fl}}$ is 858 and 3912 for ARs with flares $\geqslant$M1.0 and $\geqslant$C1.0, respectively.

\begin{figure*}[!t]
\begin{center}
\includegraphics[width=0.85\textwidth]{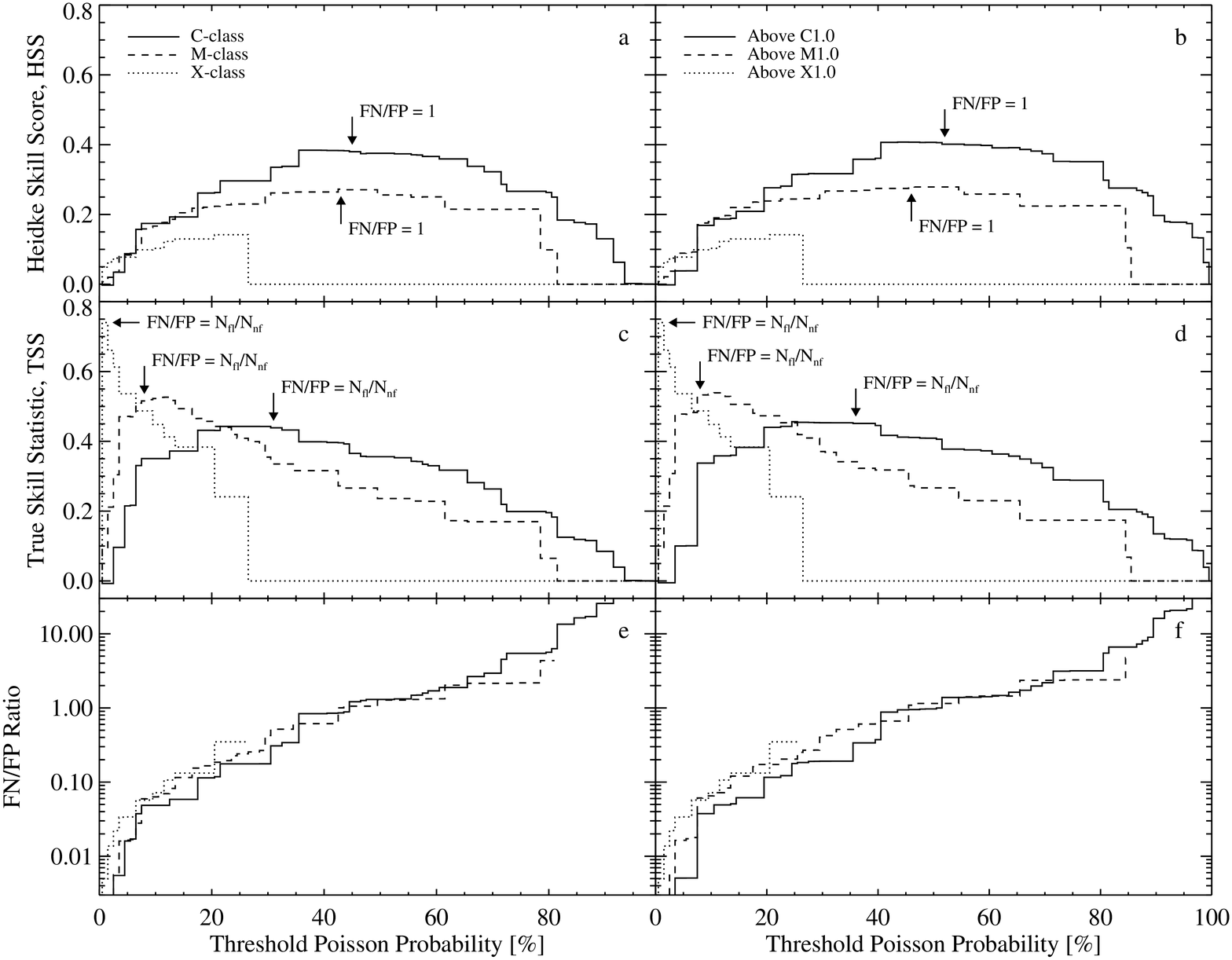}
\caption{Threshold probability variation of HSS (a-b), TSS (c-d), and FN/FP (e-f). Curves in panels (a), (c), and (e) are forecasts over 24\,hr of at least one C-class (solid), M-class (dashed), or X-class (dotted) flare, while those in panels (b), (d), and (f) are forecasts of at least one flare $\geqslant$C1.0 (solid), $\geqslant$M1.0 (dashed), and $\geqslant$X1.0 (dotted). Arrows in panels (a) and (b) mark thresholds where FN/FP $\approx$ 1, while those in panels (c) and (d) mark thresholds where FN/FP $\approx$ $N_{\mathrm{fl}}$/$N_{\mathrm{nf}}$. Only FN/FP $\approx$ $N_{\mathrm{fl}}$/$N_{\mathrm{nf}}$ is marked for X-class ($\ll$1), as 0.35 is the largest finite value.\label{fig:thr_var}}
\end{center}
\end{figure*}

The second set of information is achieved by applying a flare/no-flare discriminating threshold to the Poisson probabilities achieved in Section~\ref{subsec:pois_prob}. All ARs in the test period had the corresponding McIntosh class flare probabilities (Table~\ref{tab:classes}) assigned to the 24\,hr interval after observation. Probabilities were converted into predictions by choosing a threshold (varying in 1\% increments from 0\% to 100\%) and predicting ``no flare'' for values below the threshold and ``flare'' for those at or above the threshold.

The contingency table elements (Section~\ref{sec:int} and Table~\ref{tab:ex_con}) are the number of each pair combination of flare truth and prediction. The variation of the HSS and TSS measures are shown in Figure~\ref{fig:thr_var} and Table~\ref{tab:scores} for separate forecasts of C-, M-, and X-class events and forecasts $\geqslant$M1.0 and $\geqslant$C1.0. It is worth noting that the approach applied here changes occurrences of TP to FN and FP to TN as the threshold probability rises (``flare'' predictions become ``no flare'' predictions, but flare truth is unchanged).

\begin{deluxetable}{lrrrrccc}[!h]
\tabletypesize{\scriptsize}
\tablecaption{Flare Forecast Contingency Table and Skill Score Dependence on Threshold Poisson Probability\label{tab:scores}}
\tablewidth{\columnwidth}
\tablehead{
\colhead{Prob.}    &  \multicolumn{7}{c}{Flaring In \emph{GOES} M-class Within 24\,hr}  \\
  &  \multicolumn{4}{c}{Contingency Table Elements}  &  \multicolumn{2}{c}{Skill Scores}  &  \colhead{FN/FP}  \\
\colhead{\%}  &  \colhead{TP}  &  \colhead{FN}  &  \colhead{FP}  &  \colhead{TN}  &  \colhead{HSS}  &  \colhead{TSS}   &  
}
\startdata
  0 &  810  &      0  &  21466  &      0  &   0.000  &   0.000  &     0.00 \\
 10 &  568  &    242  &   3832  &  17634  &   0.167  &   0.523  &     0.06  \\
 20 &  452  &    358  &   2163  &  19303  &   0.221  &   0.457  &     0.17  \\
 30 &  330  &    480  &   1129  &  20337  &   0.256  &   0.355  &     0.43  \\
 40 &  288  &    522  &    850  &  20616  &   0.264  &   0.316  &     0.61  \\
 50 &  209  &    601  &    471  &  20995  &   0.256  &   0.236  &     1.28  \\
 60 &  202  &    608  &    458  &  21008  &   0.250  &   0.228  &     1.33  \\
 70 &  149  &    661  &    308  &  21158  &   0.215  &   0.170  &     2.15  \\
 80 &   59  &    751  &    173  &  21293  &   0.099  &   0.065  &     4.34  \\
 90 &    0  &    810  &      0  &  21466  &   0.000  &   0.000  &      $\infty$  \\
 100 &    0  &    810  &      0  &  21466  &   0.000  &   0.000  &      $\infty$  
\enddata
\tablecomments{(The entire table is available online in machine-readable form.  A portion is shown for guidance regarding its form and content.)}
\end{deluxetable}

\section{Results \& Discussion}
\label{sec:res_dis}

\begin{deluxetable*}{lcccccccl}[!h]
\tabletypesize{\scriptsize}
\tablecaption{Inter-forecast Skill Score Comparison\label{tab:ss_comp}}
\tablewidth{\textwidth}
\tablehead{
\multicolumn{2}{l}{Forecast}                       &  \multicolumn{6}{c}{Verification Measure}                                                                 &  \colhead{Reference}  \\
\colhead{Flare Level}  &  \colhead{Interval (hr)}  &  \colhead{TSS}  &  \colhead{FN/FP}  &  \colhead{HSS}  & \colhead{POD}  & \colhead{FAR}  &  \colhead{ACC}  &  
}
\startdata
C-class\dotfill  &  24  &  \nodata  &  \nodata  &  0.493  &  0.772  &  0.319  &  0.811  &  \citet{Colak:2009p714}  \\
\dotfill  &  24  & 0.650  & 0.429  &  0.623  &  0.850  &  0.292  &  0.818  &  \citet{Song:2009p703}\tablenotemark{a}  \\
\dotfill  &  24  & 0.090  &  7.000  &  0.116  &  0.138  &  0.471  &  0.722  &  \citet{Yuan:2010p716}  \\
\dotfill  &  24  &  0.443  &  0.176  &  0.296  &  0.737  &  0.670  &  0.711  & This work: optimum TSS  \\
\dotfill  &  24  &  0.399  &  0.836  &  0.384  &  0.513  &  0.531  &  0.824  & This work: optimum HSS  \\
\tableline
M-class\dotfill  &  24  &  \nodata  &  \nodata  &  0.470  &  0.865  &  0.688  &  0.944  &  \citet{Colak:2009p714}  \\
\dotfill  &  24  & 0.621  & 6.000  &  0.676  &  0.647  &  0.083  &  0.873  &  \citet{Song:2009p703}\tablenotemark{a}  \\
\dotfill  &  24  & 0.054  &  1.963  &  0.061  &  0.221  &  0.643  &  0.652  &  \citet{Yuan:2010p716}  \\
\dotfill  &  24  &  0.526  &  0.070  &  0.177  &  0.693  &  0.864  &  0.829  & This work: optimum TSS  \\
\dotfill  &  24  &  0.272  &  1.002  &  0.273  &  0.299  &  0.701  &  0.949  & This work: optimum HSS  \\
\tableline
X-class\dotfill  &  24  &  \nodata  &  \nodata  &  0.169  &  0.917  &  0.967  &  0.981  &  \citet{Colak:2009p714}  \\
\dotfill  &  24  & 0.693  & 2.000  &  0.739  &  0.714  &  0.167  &  0.945  &  \citet{Song:2009p703}\tablenotemark{a}  \\
\dotfill  &  24  & 0.160  &  3.000  &  0.205  &  0.206  &  0.562  &  0.843  &  \citet{Yuan:2010p716}  \\
\dotfill  &  6  & 0.312  & 0.005  &  0.008  &  0.617  &  0.992  &  0.694  &  \citet{Mason:2010p532}\tablenotemark{b}  \\
\dotfill  &  24  &  0.740  &  0.005  &  0.049  &  0.859  &  0.971  &  0.881  & This work: optimum TSS  \\
\dotfill  &  24  &  0.241  &  0.348  &  0.142  &  0.250  &  0.896  &  0.988  & This work: optimum HSS  \\
\tableline
$\geqslant$M1.0\dotfill  &  24  & \nodata  &  \nodata  &  0.153  &  \nodata  &  \nodata  &  0.922  &  \citet{Barnes:2008p206}\tablenotemark{c}  \\
\dotfill &  48  & 0.650  &  1.105  &  0.650  &  0.817  &  0.169  &  0.825  &  \citet{Yu:2009p49}\tablenotemark{d}  \\
\dotfill &  48  &  $\sim$0.66  & \nodata  &  $\sim$0.66  &  $\sim$0.90  &  \nodata  &  \nodata  &  \citet{Huang:2010p717}  \\
\dotfill  &  24  &  0.539  &  0.072  &  0.190  &  0.704  &  0.854  &  0.830  & This work: optimum TSS  \\
\dotfill  &  24  &  0.273  &  1.089  &  0.280  &  0.298  &  0.684  &  0.948  & This work: optimum HSS  \\
\tableline
$\geqslant$C1.0\dotfill  &  24  &  \nodata  &  \nodata  &  0.512  &  0.814  &  0.301  &  0.805  &  \citet{Colak:2009p714}  \\
\dotfill  &  24  & 0.641   & 0.952  &  0.636  &  0.662  &  0.349  &  0.961  &  \citet{Ahmed:2012}\tablenotemark{e}\tablenotemark{f}  \\
\dotfill  &  24  &  0.456  &  0.178  &  0.315  &  0.753  &  0.649  &  0.712  & This work: optimum TSS  \\
\dotfill  &  24  &  0.412  &  0.942  &  0.407  &  0.520  &  0.495  &  0.826  & This work: optimum HSS  
\enddata
\tablenotetext{a}{Model (4).}
\tablenotetext{b}{Reported HSS contains miscalculation of expected correct random forecasts (J.P. Mason 2011, private communication).}
\tablenotetext{c}{Total unsigned magnetic flux.}
\tablenotetext{d}{Contingency table provided by X. Huang (2011, private communication).}
\tablenotetext{e}{Temporally segmented training and operational testing (test still spatially segmented to ARs $\leqslant$60$^\circ$ from disk centre).}
\tablenotetext{f}{Contingency table calculated from reported forecast measures.}
\end{deluxetable*}

\subsection{Skill Score Variation With Prediction Threshold}
\label{subsec:thr_var}
Figure~\ref{fig:thr_var} shows HSS peaking at FN/FP$\approx$1 (panels (\ref{fig:thr_var}a) and (\ref{fig:thr_var}b)). This indicates that the HSS measure of forecast accuracy is maximized\footnote{The concept of a peak value of skill score is only possible here because forecast performance is altered by varying the threshold. Methods without a variable threshold can only achieve one value.} when the \emph{absolute} frequency of incorrect predictions are equal, FN $\approx$ FP. Sensitivity to FN/FP confirms the HSS dependence on sample ratio \citep[Equation~\ref{eqn:hss_new} here;][]{Woodcock:1976p384}. Table~\ref{tab:ex_con} shows that TP and FN increase if additional flaring ARs are included (FN/FP increases and unity occurs at higher thresholds). Conversely, FP and TN increase if additional no-flare ARs are included (FN/FP decreases and unity occurs at lower thresholds). Note that varying the number of ARs included in the verification test does not have the same effect as varying the threshold used to construct the contingency tables: adding ARs alters the sample ratio, but maintains the forecast success ratio (if the added sample is random); varying the threshold maintains the sample ratio, but alters the forecast success ratio.

Figure~\ref{fig:thr_var} also shows TSS peaking at FN/FP $\approx$ $N_{\mathrm{fl}}$/$N_{\mathrm{nf}}$ (panels (\ref{fig:thr_var}c) and (\ref{fig:thr_var}d)), where $N_{\mathrm{nf}}$ is the number of no-flare ARs ($N_{\mathrm{nf}}=22276-N_{\mathrm{fl}}$). This indicates that the TSS measure of accuracy is maximized when the \emph{fractional} frequency of incorrect predictions for flare ARs equals the \emph{fractional} frequency of incorrect predictions for no-flare ARs, FN/$N_{\mathrm{fl}}$ = FP/$N_{\mathrm{nf}}$. This dependence on the \emph{fractional} form of incorrect frequencies again illustrates that forecasts with differing sample ratios will keep the same TSS value: changes in FN or FP are absorbed by corresponding changes in $N_{\mathrm{fl}}$ or $N_{\mathrm{nf}}$ (Equation~\ref{eqn:tss_new}). Note that HSS = TSS when $N_{\mathrm{fl}}=N_{\mathrm{nf}}$, but this is seldom the case in flare forecasting as flares are rare events.

\subsection{Inter-forecast Skill Score Comparison}
\label{subsec:opt_ss_com}
Flare forecasting studies do not usually quote values of TSS and rarely use equal flare/no-flare sample sizes that make HSS equal TSS.\footnote{This behaviour is good practice given the rarity of flare events. Forcing a balance between $N_{\mathrm{nf}}$ and $N_{\mathrm{fl}}$ results in discarding $\sim$80\%, $\sim$96\%, and $>$99\% of the available $N_{\mathrm{nf}}$ sample when considering events $\geqslant$C1.0, $\geqslant$M1.0, and $\geqslant$X1.0, respectively.} Unfortunately, most do not show contingency tables that would enable TSS or other unpublished measures to be calculated. Optimum values of TSS and HSS achieved by Poisson probabilities in Section~\ref{subsec:thr_var} are compared to other methods in Table~\ref{tab:ss_comp}, restricted to those with a contingency table (or values one can be inferred from) and those quoting HSS. Other measures used in flare forecasting include the probability of detection: $\mathrm{POD}=\mathrm{TP}/[\mathrm{TP}+\mathrm{FN}]$; the false alarm ratio: $\mathrm{FAR}=\mathrm{FP}/[\mathrm{TP}+\mathrm{FP}]$; and the odds ratio or accuracy: $\mathrm{ACC}=[\mathrm{TP}+\mathrm{TN}]/[\mathrm{TP}+\mathrm{FN}+\mathrm{FP}+\mathrm{TN}]$. Table~\ref{tab:ss_comp} includes these to allow broad assessment of each method.

\subsubsection{Performance for Separate Flare-magnitude Classes}
\label{subsubsec:ind_cla}
In forecasting flares in the separate \emph{GOES} flare classes over 24\,hr intervals, the ordinal logistic regression model (4) of \citet{Song:2009p703} yields the highest TSS values for C- and M-classes, while the optimum TSS for Poisson probabilities is highest for X-class. \citet{Song:2009p703} convert flare probabilities into predictions using static thresholds of 50\% for C- and M-class events and 25\% for X-class events. Improved performance might be achieved by the \citet{Song:2009p703} technique by investigating its dependence on the prediction threshold, as studied here. Unfortunately, the \citet{Song:2009p703} results are the most susceptible to noise (given a small sample of 55 ARs\footnote{Changing 1 TP into FN (and vice versa) yields $\pm$0.050, $\pm$0.059, and $\pm$0.143 in TSS for C-, M-, and X-class forecasts, respectively.}) and weighted toward successful prediction of flaring ARs, since their samples of each flare-magnitude class have higher proportions of flaring ARs (36\%, 31\%, and 13\% for C-, M-, and X-classes) than typically observed (16\%, 4\%, and 0.4\% in cycle 23). It is unclear how this method would perform operationally when non-flaring ARs outnumber flaring ARs and successfully predicting no-flare periods has increased importance. The significantly lower performance of \citet{Yuan:2010p716} in TSS and HSS is surprising with adding support vector machine classification to the \citet{Song:2009p703} technique. It is worth noting that neural network operational forecasting of McIntosh classes by \citet{Colak:2009p714} yields an HSS between that found here and \citet{Song:2009p703} for all flare classes, but published values do not permit TSS calculation.

For X-class flares, the optimal TSS value for Poisson probabilities over 24\,hr intervals is higher than that from the superposed-epoch analysis of \citet{Mason:2010p532} over 6\,hr intervals. The \citet{Mason:2010p532} technique is segmented by predicting ``no flare'' for ARs with a magnetic quantity change over the previous 40\,hr below one threshold and ``flare'' for ARs with changes above a second higher threshold. The forecast success would likely decrease if the unpredicted mid-range AR population were included. Note the optimum TSS found here has large FAR because it results from a yes/no prediction threshold of 1\%, meaning that X-class flares are always predicted for all McIntosh classifications that historically produced any X-class activity.

\subsubsection{Performance above the M1.0 Level}
\label{subsubsec:gtr_M1}
In forecasting flares $\geqslant$M1.0, sequential supervised learning by \citet{Yu:2009p49} and the predictor team work of \citet{Huang:2010p717} yield the highest HSS values that equate to TSS from equal flare and no-flare sample sizes. However, they predict cumulative flare importance equivalent to at least one M1.0 event in a 48~hr interval (e.g., 10 C1.0, 5 C2.0, 2 C5.0). This raises uncertainty about these good skill scores representing the successful forecasting of events $\geqslant$M1.0, as forecasting multiple C-class events from an AR may be easier than single M-class events. More importantly, both works only consider ARs that produce at least one flare $\geqslant$C1.0 in their life. This segmentation weakens their interpretation for operational purposes (similar to the case of \citet{Song:2009p703} in Section~\ref{subsubsec:ind_cla}), as the number of AR no-flare periods considered in \citet{Yu:2009p49} and \citet{Huang:2010p717} are severely reduced by excluding all completely non-flaring NOAA numbers. It is worth noting that the optimum TSS achieved here equals that for the application of 1 decision tree in \citet{Huang:2010p717} (with HSS, hence TSS, of $\sim$0.54).

The highest HSS achieved in the discriminant analysis study of \citet{Barnes:2008p206} was found using total unsigned magnetic flux. However, the value is low (notably also lower than the optimum HSS found here) and likely due to the overlap between flaring and non-flaring AR-parameter distributions. However, proper comparison to the performance of Poisson probabilities is not possible as TSS values from \citet{Barnes:2008p206} are not available.

\subsubsection{Performance above the C1.0 Level}
\label{subsubsec:gtr_C1}
Finally, in forecasting flares $\geqslant$C1.0 in 24~hr intervals, the application of neural networks by \citet{Ahmed:2012} to magnetic properties with semi-operational testing yields the highest TSS. Semi-operational refers to no segmentation being applied based on flare history, while spatial segmentation was applied (only ARs within 60$^\circ$ of disk centre). Optimum TSS values show that Poisson probabilities do not perform as well as the machine learning of \citet{Ahmed:2012}, possibly from truly operational application (e.g., ARs near the limb may be misclassified by foreshortening effects and inappropriately predicted). It is interesting that the neural network system of \citet{Colak:2009p714} does not perform significantly better than the application of Poisson probabilities, but this is based on HSS as TSS is unavailable for their work.

\section{Conclusions}
\label{sec:con}
To be operationally practical, flare forecasts should provide predictions for all ARs irrespective of properties or flare history (i.e., no minimum criteria in selecting ARs for flare prediction). We have presented the variation of forecast verification measures with the threshold Poisson probability used to define ``flare'' and ``no flare'' predictions. Forecasts for different X-ray flare levels from all NOAA ARs over 1996 August 1 to 2010 December 31 were tested against observed flares.

Optimized forecasts from Poisson flare probabilities are found to perform to similar standards as some more sophisticated methods (e.g., in forecasting events $\geqslant$M1.0). However, the relatively low levels of optimum skill score (HSS  $\lesssim$ 0.4 and TSS\footnote{Optimum TSS of 0.74 is found here for X-class at a threshold of 1\%, but this results in severe overprediction and large FAR.} $\lesssim$ 0.5) lend further support to the need to use flaring rate distributions \citep[in, e.g., a Bayesian methodology like][]{Wheatland:2005p71} rather than averages over an AR class. This will be a focus of future work in the construction of Bayesian prior distributions of AR-property-dependent flare rates.

Providing forecasts and quantifying their performance will be acutely necessary as we approach the activity maximum of cycle 24. It is foreseen that specific forecast requirements may be targeted by careful consideration of skill scores and particular contingency table elements, e.g., the threshold for interpreting flare probabilities as yes/no forecasts could be tailored to achieve relative failure ratios (FN/FP) within the tolerance of various groups in the scientific and space weather communities. However, complete flare forecasts will require a deeper physical understanding of magnetic energy release and partitioning of energy between flare emission at different temperatures, acceleration of CMEs, and acceleration of high-energy particles \citep{Emslie:2005}.
 
In closing, it is imperative that the performance of flare forecasting methods with differing flare/no-flare sample ratios is compared in a suitable manner. This requires the use of a verification measure that is not sensitive to the flare/no-flare sample ratio. We have highlighted an issue with the commonly adopted HSS and instead propose the sample ratio invariant TSS for the reliable comparison of flare forecasts.

\acknowledgments
The authors thank the referee for comments that improved the paper, Christopher Balch (NOAA/SWPC) for providing flare-AR-association data over 1988--1996, the NASA All-clear Workshop 2008, and Graham Barnes for useful discussions. This work was supported by a Marie Curie Intra-European Fellowship (D.S.B.) and the HELIO e-Infrastructure grant (P.A.H.) under the European Community's 7th Framework Programme.

{\it Facilities:} \facility{\emph{GOES} (XRS)}


\end{document}